\documentclass[twocolumn,showpacs,amsmath,amssymb]{revtex4}

\usepackage{graphicx}
\usepackage{bm}

\begin{document}


\title{Reliability of temporal coding on pulse-coupled networks of oscillators}
\author{Jun-nosuke Teramae}
\email{teramae@brain.riken.jp}
\author{Tomoki Fukai}
\affiliation{Laboratory for Neural Circuit Theory, RIKEN Brain Science Institute, Saitama, Japan}


\begin{abstract}
 We study the reliability of spike output in a general class of
 pulse-coupled oscillators receiving a fluctuating input. Showing that
 this problem is equivalent to noise-induced synchronization between
 identical networks of oscillators, we employ the phase reduction method
 to analytically derive the average Lyapunov exponent of the
 synchronized state. We show that a transition occurs between reliable
 and unreliable responses at a critical coupling strength, which is
 determined through the competition between the external input and
 recurrent input. To our surprise, the critical value does not depend on
 intrinsic properties of oscillators. 
\end{abstract}

\pacs{05.45.Xt, 02.50.Ey, 05.10.Gg, 87.18.Sn}


\maketitle

Noise-induced synchronization appears in a variety of phenomena
including lasers \cite{laser}, chemical reactions \cite{chemical},
gene networks \cite{gene} and 
neuronal systems \cite{mainen,neuron}. In these systems, periodic or
chaotic oscillators 
driven by a common fluctuating input synchronize with each other due to
the nonlinearity of oscillators and the stochastic nature of the input
\cite{nis}. The phase reduction \cite{kuramoto} and the Lyapunov
analysis proved that two or
more identical oscillators receiving a common fluctuating input are
always in-phase synchronized regardless of their intrinsic properties
and initial phases \cite{teramae,piko_nakao}. We can interpret such
oscillators as a single 
oscillator receiving the same input repeatedly, but with different
initial phases, i.e. many trials of an input application. Therefore, the
in-phase synchronization of input-driven oscillators implies, in a
single oscillator, the reproducibility of the responses to a repeated
input, or response reliability, which is particularly important for
processing external signals. Reliable responses to a fluctuating input
are actually measured from single cortical neurons
\cite{mainen}. However, neurons and other oscillators in the real
world work collectively in their networks rather than individually. To
study whether a network of oscillators still has response reliability,
we develop a theory of noise-induced synchronization between networks of
oscillators rather than between single oscillators. We find a transition
from reliable to unreliable responses at a critical coupling
strength. Deriving average Lyapunov exponent analytically, we reveal
that the critical value is determined through the competition between
variance of the external input and of internal recurrent inputs
regardless of details of oscillators. Around the transition
point where magnitude of the average Lyapunov exponent is small,
information of initial states can stay in the network for a long
time. We discuss a possible role of the long time scale in role-sharing
between rate and temporal coding on neuronal computation in the brain.

\begin{figure}
 \includegraphics{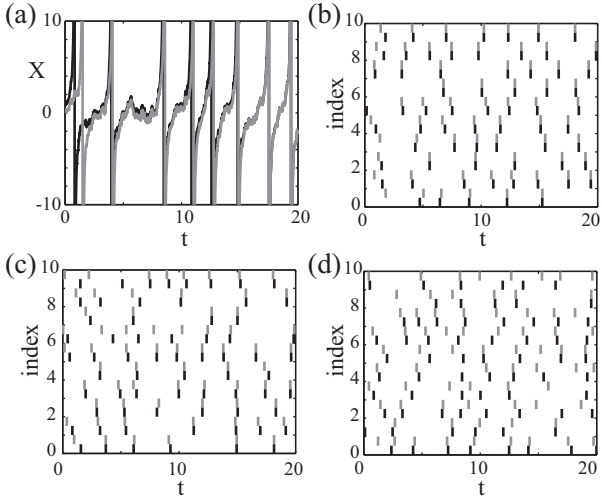}
 \caption{
 Responses reliability of a pulse-coupled network of quadratic
 integrate-and-fire models. $I=1.0$, $\sigma=1.0$, $N_E=80$ and
 $N_I=20$ (a) Time evolutions of $X$ of an oscillator for 1st (black)
 and 2nd (gray) trials when
 $g=0$. (b-d) Raster plots of spike times of randomly chosen 10
 oscillators of the network for 1st (black) and 2nd trials
 (gray). $g=0$ (b), $0.1$ (c) and $0.3$ (d).
 }
 \label{figure:1}
\end{figure}

A network of pulse-coupled $N$ limit-cycle oscillators receiving
fluctuating inputs are described as:
\begin{equation}
 \frac{d\bm{X}_i}{dt}=\bm{F}(\bm{X}_i)+\bm{\xi}_{i}(t)+\sum_{j=1}^N\sum_{spike}
  g_{ij}\delta(t-t_j^{spike})\hat{\bm{g}},
 \label{eq:1}
\end{equation}
where $i=1,\cdots ,N$ and $\dot{\bm{X}}=\bm{F}\left(\bm{X}\right)$ has a
stable limit-cycle solution $\bm{X}_0\left(t\right)$. A unit vector
$\hat{\bm{g}}$ indicates the direction of interactions in the
multidimensional space spanned by $\bm{X}$. We assume, with neuronal
oscillators in mind, coupling matrix $g_{ij}$ is a sparse random matrix
with connection probability $p$, and each nonzero component of $g_{ij}$
is either $g$ or $-g_{I}$ if $j$ refers to an excitatory or an
inhibitory cell, respectively. The network consists of $N_{E}$
excitatory neurons and $N_{I}=N-N_{E}$ inhibitory neurons. Fluctuating
external inputs $\bm{\xi}_{i}$ represent independent white Gaussian processes
with strength $\langle \bm{\xi}_{i}\left(t\right)
\bm{\xi}_{j}\left(s\right)\rangle=2\sigma^2\delta_{ij}\delta\left(t-s\right)$,
and $t_{j}^{spike}$ represents spike times of the $j$th neuron. We use
$g$ as a control parameter of the network and require, for simplicity,
that $g_{I}$ is in proportional to $g$ and satisfies the balance condition
\cite{balance,vreeswijk}, $g N_E+\left(-g_I\right) N_I=0$, whereas results of the paper are
independent of the restriction. When $g=0$, response of oscillators are
always reliable, i.e. the spike sequence of each oscillator converges
into the same sequence in different trials. Figure \ref{figure:1}a and
\ref{figure:1}b
demonstrates the reliable responses obtained from numerical calculations
of quadratic integrate-and-fire (QIF) neurons, $F\left(X\right)=X^2+I$, with
variable resetting $X\left(t^{spike}\right)=\infty \rightarrow
X\left(t^{spike}+0\right)=-\infty$ \cite{qif}. Whereas two trials start from
different
initial states, raster plots of them converge into same sequences,
i.e. same spike times. We then introduce finite couplings $g>0$
to the network of oscillators and calculate firing responses in a
similar way to Fig. \ref{figure:1}b. When coupling strength is small, the population
is still reliable, spike sequences of different trials converge into the
same one (Fig. \ref{figure:1}c). However, the reliability is lost from the population
when coupling strength is sufficiently large (Fig. \ref{figure:1}d). Spike sequences
of two trials never converge into the same one while we apply the same
input to trials. In terms of synchronization, the result means that
fluctuating inputs induce phase synchronization to two identical
networks of oscillators only when coupling strength of the network is
sufficiently weak.

Regarding the fluctuating signals and recurrent connections as
perturbations to the deterministic oscillators, we apply the standard
phase reduction method to Eq. (\ref{eq:1}) and obtain stochastic equations of
phases $\phi_{i}$ as,
\begin{equation}
 \frac{d\phi_i}{dt}=\omega+Z(\phi_i)\left(\xi_i(t)+\sum_{j=1}^N\sum_{spike} g_{ij}\delta(t-t_j^{spike})\right),
 \label{eq:2}
\end{equation}
where $\omega$ is an intrinsic frequency of the unperturbed
oscillators. Phase sensitivity function or phase response function is
defined uniquely from $\bm{X}_0(\phi)$ as
$\bm{Z}(\phi)=\rm{grad}_{\bm{X}} \phi |_{\bm{X}=\bm{X}_0(\phi)}$ \cite{kuramoto}. To
simplify notations, we assumed without loss of generality that
$\hat{\bm{g}}$ is in parallel with $\bm{\xi}$ and replaced vector
variables $\bm{Z}\left(\phi\right)\hat{\bm{g}}$ and
$\bm{Z}\left(\phi\right)\bm{\xi}$ to scalar variables
$Z\left(\phi\right)$ and $Z\left(\phi\right)\xi$ in Eq. (\ref{eq:2}). For the
QIF model, for instance, $\omega=I^{-1/2}$,
$Z\left(\phi\right)=\omega\left(1+\cos
\left(\phi\right)\right)$. Note that discontinuous variable
resetting $X\left(t^{spike}\right)=\infty \rightarrow
X\left(t^{spike}+0\right)=-\infty$ is now reduced to continuous dynamics
over a spike threshold $\phi=\pi$ because derivation of $F$ is
continuous between $X=\infty$ and $X=-\infty$,
$F'(\infty)=F'(-\infty)$. To avoid unrealistic cases where an oscillator
sends numerous spikes within a short interval of time when its
fluctuating phase crosses the firing threshold, $Z\left(\phi\right)$
should be vanish around the threshold. Realistic neuron models including
the QIF model satisfy the
condition. Reliability of the firing responses is equivalent to that of
phase dynamics because phase deviation is proportional to deviation 
of firing time. Phases of two trials, $\phi_{i}$ and $\tilde{\phi}_{i}$
evolve satisfying Eq. (\ref{eq:2}) from different initial phases but receiving
same inputs, $\xi_{i}$, in the same network, $g_{ij}$. Since phase
synchronized state, $\phi_{i}=\tilde{\phi}_{i}$, is an obvious solution
of these two equations, linear stability around the solution determines
the response reliability. To evaluate the stability we linearize Eq. (\ref{eq:2})
in terms of small phase differences,
$\psi_{i}=\tilde{\phi}_{i}-\phi_{i}$, and calculate average Lyapunov
exponent \cite{arnold} over all oscillators in the network.

Coupling terms are linearized as follows. Consider increment of
$\psi_{i}$, when $j$th cell fires. We can take firing time of
$\tilde{\phi}_{j}$ is $t=0$, firing time of $\phi_{j}$ is therefore
$t=dT=\psi_{j}/\omega$. We can assume $dT>0$ without loss of
generality. At $t=0$,  $\tilde{\phi}_{i}$ receives a spike 
from $\tilde{\phi}_{j}$,
\begin{equation}
  \begin{array}{rcl}
   \tilde{\phi}_i(0^+)&=&\tilde{\phi}_i(0)+g_{ij}Z(\tilde{\phi}_i(0))\\
   \phi_i(0^+)&=&\phi_i(0)
  \end{array}.
  \label{eq:3}
\end{equation}
Phases evolve as follows from $t=0^{+}$ to $dT$, because $dT$ is a short interval, 
\begin{equation}
  \begin{array}{rcl}
   \tilde{\phi}_i(dT)&=&
    \tilde{\phi}_i(0^+)+\left(\omega+\sigma^2 Z'Z(\tilde{\phi}_i(0^+))\right)dT\\
   & &+Z(\tilde{\phi}_i(0^+))dW\\
   \phi_i(dT)&=&\phi_i(0^+)+\left(\omega+\sigma^2 Z'Z(\phi_i(0^+))\right)dT\\
   & &+Z(\phi_i(0^+))dW
  \end{array},
  \label{eq:4}
\end{equation}
where $dW=\xi_{i}dT$. In order to evaluate phase responses $Z(\phi)$ at
precise timings just before spike inputs, we translated Eq. (\ref{eq:2}) to
equivalent Ito integrals in Eq. (\ref{eq:4}) \cite{strato}. Third terms
of Eq. (\ref{eq:4}) result from the translation. Finally, at $t=dT$,
$\phi_{i}$ receives a spike from $\phi_{j}$,
\begin{equation}
 \begin{array}{rcl}
  \tilde{\phi}_i(dT^+)&=&\tilde{\phi}_i(dT)\\
  \phi_i(dT^+)&=&\phi_i(dT)+g_{ij}Z(\phi_i(dT))
 \end{array}.
 \label{eq:5}
\end{equation}
We can linearize from Eq. (\ref{eq:3}) to (\ref{eq:5}) in terms of
$\psi$ with an attention that $dW$ is the order of $dT^{1/2}$ and then the
order of $\psi_{j}^{1/2}$. Taking all connections into account and
neglecting terms higher than the order of $g^{2}$, we obtain linearized
equation of $\psi$ as
\begin{eqnarray}
 \frac{d\psi_i}{dt} &=& \left(\sigma^2 \left(Z'(\phi_i) Z(\phi_i)\right)'+Z'(\phi_i)
	       \xi_i\right)\psi_i  \label{eq:6}\\
 & &+\sum_{j=1}^N\sum_{spike} g_{ij} Z'(\phi_i)
 \delta (t-t_j^{spike}) \left(\psi_i-\psi_j\right). \nonumber
\end{eqnarray}
By introducing new variables $y_{i}=\left(\log{\psi_{i}^{2}}\right)/2$,
Eq. (\ref{eq:6}) is further rewritten as
\begin{eqnarray}
 \frac{d y_i}{dt} &=& \sigma^2 \left(Z'(\phi_i)
				Z(\phi_i)\right)'+Z'(\phi_i)
 \xi_i-\sigma^2 Z'(\phi_i)^2 \label{eq:7} \\
 & &+\sum_{j=1}^N\sum_{spike} \delta (t-t_j^{spike}) \log
 \left| 1+g_{ij} Z'(\phi_i)\left(1-\frac{\psi_j}{\psi_i}\right)\right|.\nonumber
\end{eqnarray}
Since the Lyapunov exponent $\lambda_{i}$ is defined as
$\lim_{T\to\infty}(y_{i}(T)-y_{i}(0))/T$, the long time average of the
Eq. (\ref{eq:7}) coincides with $\lambda_{i}$. We assume that the network is in
asynchronous steady firing state due to fluctuating inputs and replace
spike times of cells by independent Poisson processes with firing rate
$r$ \cite{asynchro}. Then averaging of Eq. (\ref{eq:7}) over the Poisson
processes and over all oscillators in the network gives
\begin{equation}
 \langle \frac{d y_i}{dt} \rangle
  =\langle Z'^2 \rangle \left(
			 -\sigma^2+Nr \frac{\langle g^2 \rangle}{2}
			 \left(
			  1+\langle \frac{\psi_j^2}{\psi_i^2} \rangle
			 \right)
			\right),
  \label{eq:8}
\end{equation}
where $\langle Z'^2 \rangle=(2\pi)^{-1}\int_{0}^{2\pi} Z'\left(\phi\right)^2
d\phi$. Here we used the assumption of weak inputs and
weak interactions and
reduced distributions of phases to uniform distributions in
$[0,2\pi]$. Unfortunately, Eq. (\ref{eq:8}) is not a closed form of
$y_{i}$. However, when variance of $\psi^{2}$ is small, or $\langle
\psi^{4} \rangle\simeq \langle \psi^{2} \rangle^{2}$, the last term of
Eq. (\ref{eq:8}) is approximated as $\langle \psi_{i}^{2}/\psi_{j}^{2} \rangle=1$
and we finally obtain the following main formula of the average Lyapunov
exponent:
\begin{equation}
 \lambda=\langle Z'^2 \rangle \left(-\sigma^2+Nr\langle g^2 \rangle\right).
  \label{eq:9}
\end{equation}
Note that when variance of $\psi^{2}$ is not small, Eq. (\ref{eq:9}) gives the
lower bound of $\lambda$ because $\langle
\psi_{i}^{2}/\psi_{j}^{2} \rangle\geq 1$ in generally.

To confirm the above analysis, we calculate averaged dynamics of
$y_{i}=\left(\log{\psi_{i}^{2}}\right)/2$ numerically for networks of
QIF oscillators. Due to fluctuating inputs and
recurrent interactions, $y_{i}$ themselves do not evolve
monotonically. However, population averages of $y_{i}$ decrease or
increase almost linearly depending on coupling strengths g as predicted
by Eq. (\ref{eq:9}).
\begin{figure}
 \includegraphics{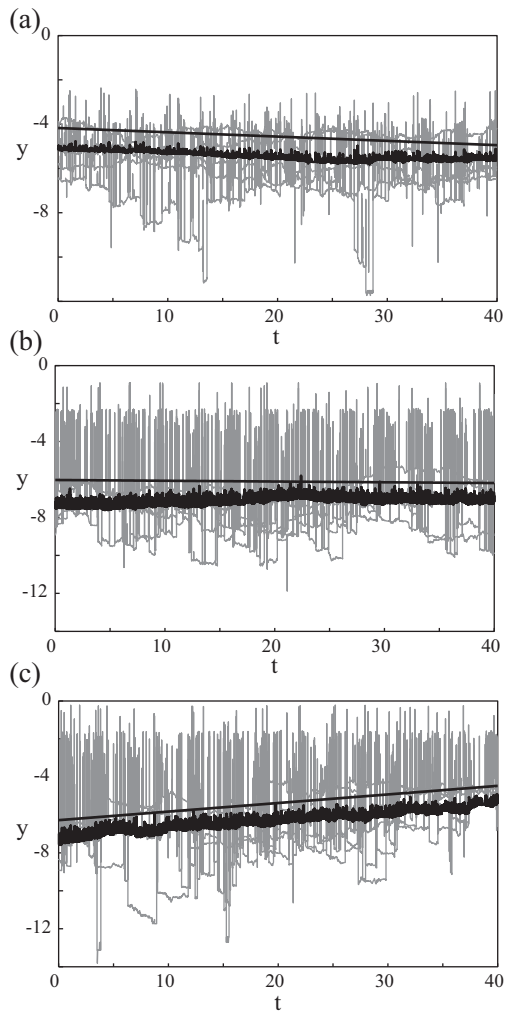}
 \caption{
 Time evolutions of logarithm of phase differences $y_i$ of randomly chosen 5
 oscillators (gray lines). Sudden tentative jumps to large values on these lines
 are because of differences of spike times between two trials (see
 Eq. (\ref{eq:3}) and (\ref{eq:5})). Thick black curves are averages of
 $y_{i}$ over all oscillators in the
 network. Slopes of these curves agree well with the
 analytical results of Eq. (\ref{eq:9}) shown by slopes of thin black lines. The
 values of parameters are the same as in Fig. \ref{figure:1} except that $\sigma=0.2$,
 and $g=0.01$ (a), $0.05$ (b) and $0.1$ (c).
 }
 \label{figure:2}
\end{figure}

Our expression of the Lyapunov exponent, Eq. (\ref{eq:9}), tells us two important
facts of the reliability transition. First, the transition stems from a
competition between two variances, variance of input signals $\sigma^2$
and variance of recurrent inputs $Nr \langle g^{2} \rangle$. Whereas the
first contribution to the exponent is negative, the second is always
positive. Therefore, the network lost their reliability when the second
exceeds the first. Second, the factor of $\langle Z'^{2} \rangle$ which
reflects intrinsic properties of oscillators is multiplied equivalently
to these two factors $\sigma^2$ and $Nr \langle g^{2} \rangle$ in
Eq. (\ref{eq:9}). Therefore, the critical coupling strength $g_{c}$, given as the
solution of $\lambda=0$, is universal in the sense that $g_{c}$ is
independent of details of oscillators. For the network structure we used
in Fig. \ref{figure:1}, the critical value is given as $g_{c}=\sigma \sqrt{(Npr)^{-1}
N_{I}/N_{E}}$ regardless of oscillators on the network. If we use
another natural normalization of coupling strengths as $g\rightarrow
g/\sqrt{Np}$ \cite{vreeswijk}, we can eliminate $N$ from the critical strength,
$g_{c}=\sigma \sqrt{r^{-1} N_{I}/N_{E}}$.

In the vicinity of the critical coupling strength $g=g_{c}$ where
$\left|\lambda\right|\ll 1$, information on the initial states of
oscillators may disappear quite slowly after the onset of input. This
slow transient behavior might have the following implications for
computations by cortical networks. The output of the computation is not
simply determined by the current input, but is also modulated by the
brain's internal state and/or input histories \cite{history,liquid}. In our model, the
membrane time constant sets the short time scale that enables the
network to respond quickly to an external input with firing rate of
population dynamics \cite{vreeswijk,rate}. By contrast, the critical dynamics of temporal
spike sequences may set a much longer time scale to ensure the response
diversity reflecting the initial state or input histories. This
implies that neuronal populations may simultaneously achieve two
different time scales by parallel use of rate code and temporal
code. Further studies are required for clarifying this possibility.

\begin{figure}
 \includegraphics{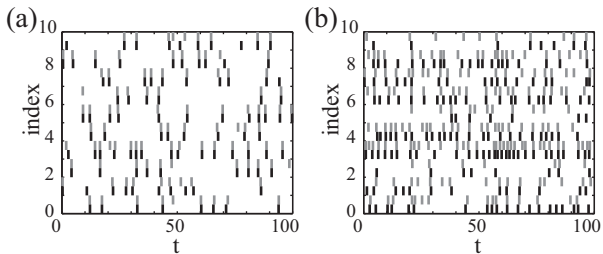}
 \caption{
 Responses reliability of a network of sub-threshold QIF
 models for two trials. The values of parameters are the same as in
 Fig. \ref{figure:1} except that
 $I=-1.0$, and $g=0.4$ (a) and $1.0$ (b).
 }
 \label{figure:3}
\end{figure}

So far, we have restricted our study to super-threshold neurons which
continue to fire without external inputs. Numerical simulations of
sub-threshold QIF model with $I<0$, however,
suggest that similar transition also occurs in a network of
sub-threshold neuron models (figure \ref{figure:3}). It remains unknown
whether this transition
may appear in a broad class of sub-threshold neuron models because the
phase reduction method is not applicable to sub-threshold neuron
models. A unified treatment of super- and sub- threshold neuron models
is awaited. We have assumed that couplings among oscillators are
delta-functions. To remove a doubt that our results might be
pathological phenomena come from singularity of delta-functions, we
calculated reliability of coupled oscillators numerically using a
alpha-function, $\alpha(t)=\alpha^{2}t\exp(-\alpha t)$, instead of
$\delta(t)$. Again, we could see similar transition from reliable to
unreliable responses (results not shown). Linear integrate-and-fire
model is the most useful description of firing neurons. This model,
however, behave unrealistically about response reliability even when
$g=0$ because of its anomalous variable resetting \cite{teramae,lif}. Here, we use
quadratic integrate-and-fire model to avoid the problem. Coupled
oscillators may synchronize with each other if $g$ is sufficiently large,
whereas we have only concentrated on the asynchronous steady
state. Synchronization may affect average Lyapunov exponent and
may change the transition significantly because we must use correlated
stochastic processes instead of independent Poisson processes when we
average Eq. (\ref{eq:8}) to obtain $\lambda$.

In conclusion, coupled reliable elements are not necessarily reliable
any more. Spike responses of coupled oscillators to fluctuating inputs
show transition from reliable responses to unreliable responses.
In terms of noise-induced synchronization, common noises fail
to induce phase synchronization to networks of strongly coupled
oscillators whereas same inputs always induce synchronization to single
oscillators. Underlying mechanism of the transition is competition
between a variance of external signals and a variance of internal
recurrent inputs. Critical coupling strength derived analytically is
independent of details of oscillators because phase response functions
appear equivalently in these competing factors.

We thank Y. Tsubo and H. Cateau for fruitful discussions and valuable
comments. The present work was supported by Grant-in-Aid for Young
Scientists (B) 50384722 and Grant-in-Aid for Scientific Research
17022036 from the Japanese Ministry of Education, Culture, Sports,
Science and Technology.

\end{document}